\documentstyle[prb,aps,floats,epsfig]{revtex}

\begin{document}
\draft

\twocolumn[\hsize\textwidth\columnwidth\hsize\csname @twocolumnfalse\endcsname

\title{Exchange-correlation kernels for excited states in solids}
\author{Krzysztof Tatarczyk\cite{email}, Arno Schindlmayr, and Matthias
  Scheffler}
\address{Fritz-Haber-Institut der Max-Planck-Gesellschaft, Faradayweg 4--6,
  14195 Berlin-Dahlem, Germany}
\date{\today}
\maketitle

\begin{abstract}
The performance of several common approximations for the exchange-correlation
kernel within time-dependent density-functional theory is tested for
elementary excitations in the homogeneous electron gas. Although the adiabatic
local-density approximation gives a reasonably good account of the plasmon
dispersion, systematic errors are pointed out and traced to the neglect of the
wavevector dependence. Kernels optimized for atoms are found to perform poorly
in extended systems due to an incorrect behavior in the long-wavelength limit,
leading to quantitative deviations that significantly exceed the experimental
error bars for the plasmon dispersion in the alkali metals. {\em Copyright
  2000 by The American Physical Society.}
\end{abstract}

]

\section{Introduction}

Exchange and correlation effects are crucial for understanding the properties
of interacting many-electron systems but notoriously difficult to implement
accurately in {\em ab initio\/} computational schemes. In the Kohn-Sham
formulation of density-functional theory\cite{Hohenberg1964} they are
incorporated into the exchange-correlation potential, which is a functional of
the electron density but in practice is not known exactly except for simple
model systems. Hence approximations such as the local-density or generalized
gradient approximations are needed. In many cases these yield accurate total
energies and related ground-state quantities.\cite{Dreizler1990} However, the
development of experimental devices that allow, in principle, to track a
single electron and the emergence of new fields such as surface photochemistry
place increasing emphasis on the study of {\em excited\/}
states. Unfortunately its variational foundation prevents a straightforward
application of the Kohn-Sham scheme to electronic excitations, for which
numerically expensive Green function techniques or, in case of small systems,
quantum-chemical methods were traditionally employed.

Time-dependent density-functional theory\cite{Runge1984} promises an
appealing alternative. Originally designed to explore time-dependent
phenomena, it was recently realized that it can also be exploited to
investigate optical excitations, which involve the creation of electron-hole
pairs.\cite{Petersilka1996} These are to be distinguished from photoemission
states, which can be determined by calculating the self-energy correction to
the quasiparticle band structure, for instance in the $GW$
approximation.\cite{Aryasetiawan1998} The optical excitation energies
correspond to the poles of the full linear density-response function, which
can, in principle, be obtained exactly within time-dependent
density-functional theory. This approach has since been applied to excited
states not only in the context of quantum chemistry\cite{Petersilka1996,%
Casida1998a,Hirata1999,Sundholm1999} but also, more recently, in solid-state
physics.\cite{Olevano1999,Kootstra2000,Quong1993,Ku1999,Cazalilla2000} The
procedure starts from time-independent ground-state properties, such as the
Kohn-Sham orbitals and eigenvalues, which are conveniently obtained by
conventional means. However, besides static exchange and correlation embodied
in the exchange-correlation potential, {\em dynamic\/} effects due to the
time-dependent perturbation must also be accounted for. The latter are
described by the so-called exchange-correlation kernel. Formally the kernel is
a functional derivative of the exchange-correlation potential, evaluated at
the unperturbed ground-state density, but as the exact potential is unknown
and must in practice be approximated by a parameterization, independent
expressions are often used for both quantities.

Chemical studies have repeatedly emphasized that an accurate description of
the exchange-correlation potential is particularly important for excited-state
calculations of atoms and molecules.\cite{Umrigar1994,Casida1998b,Burke2000}
Although the local-density and generalized gradient approximations often yield
good total energies, the corresponding potentials fail to capture the correct
Coulomb-like asymptotic behavior and instead decay exponentially. As a result,
many unoccupied states that really should be bound are pushed to higher
energies and merge with the continuum. The accessible excitation spectrum is
then poorly rendered. Much effort is therefore invested into better
expressions for the exchange-correlation potential. In comparison, the kernel
is considered to be less important, and the adiabatic local-density
approximation is often chosen for convenience.

It is not clear whether these findings apply equally to solids. First, the
problem of possible unbound states becomes irrelevant in bulk
materials. Second, nonlocal dynamic exchange and correlation effects, which
are neglected in the adiabatic local-density approximation, naturally become
more prominent in extended systems with delocalized wavefunctions. For solids
the errors introduced by approximations for the potential and the kernel could
therefore be of similar magnitude. However, although several new
parameterizations for the kernel were recently proposed in the
literature,\cite{Petersilka1996,Burke2000,Corradini1998} these have not yet
been systematically applied to solids, despite first encouraging
attempts.\cite{Olevano1999} In fact, almost all calculations reported so far,
which focus either on the dielectric response of
semiconductors\cite{Olevano1999,Kootstra2000} or the plasmon dispersion in
simple\cite{Quong1993,Ku1999} and noble\cite{Cazalilla2000} metals, employ the
local-density approximation both in the potential and the kernel. In contrast
to atoms, this approach seems to work reasonably well for solids, but
improvements are still desirable.

In order to understand and quantify the error introduced by approximations to
the kernel, more detailed studies for solids are necessary. Lein and
co-workers\cite{Lein2000} have recently compared the correlation energy of the
homogeneous electron gas for different parameterizations, but the energy is an
integrated quantity that principally reflects average weight distributions and
is less sensitive to variations in the small-scale structure of the
kernel. However, the latter have a significant influence on the excitation
spectrum, which is given by the position of poles in the linear
density-response function. In this paper we therefore concentrate on the
dispersion relations for the plasmon frequency and lifetime, assessing the
performance of several kernels currently circulated in the literature. By
applying them to the homogeneous electron gas, where the exchange-correlation
potential is a trivial constant, we are able to isolate the error due to the
kernel and make a systematic comparison. Indeed, we find that the choice of
parameterization plays an important role, and that inappropriate kernels
optimized for atoms give rise to quantitative deviations that significantly
exceed the experimental error bars for the plasmon dispersion in the alkali
metals. We believe that the findings presented in this paper can not only be
extended to plasmons in real materials but generally apply to excited states
in solids that, like plasmons, are based on charge rearrangements. A prominent
example are charge-transfer excitations in surface-adsorbate systems that
occur during photoinduced reactions.\cite{Kluner1998}

This paper is organized as follows. In Sec.\ \ref{Sec:Method} we give an
outline of our computational method. The kernels considered here are listed in
Sec.\ \ref{Sec:Kernels}, and in Sec.\ \ref{Sec:Results} we present the
numerical results together with a discussion. Finally, in Sec.\
\ref{Sec:Summary} we summarize our conclusions. Rydberg atomic units are used
throughout.

\section{Computational method}\label{Sec:Method}

Within linear response theory the true many-body density-density response
function is defined as
\begin{equation}
\chi({\bf r},{\bf r}';t-t') = \frac{\delta n({\bf r},t)}{\delta V^{\rm
    ext}({\bf r}',t')} \;,
\end{equation}
where $\delta n({\bf r},t)$ indicates the density change induced by an
external perturbation $\delta V^{\rm ext}({\bf r'},t')$, and the functional
derivative is evaluated at the static external potential corresponding to the
unperturbed ground-state density. Likewise, the Kohn-Sham susceptibility
\begin{equation}
\chi^0({\bf r},{\bf r}';t-t') = \frac{\delta n({\bf r},t)}{\delta V^{\rm
    eff}({\bf r}',t')} \;,
\end{equation}
describes the response of the associated noninteracting system with the same
electron density due to a change $\delta V^{\rm eff}({\bf r}',t') = \delta
V^{\rm ext}({\bf r'},t') + \delta V^H({\bf r}',t') + \delta V^{\rm xc}({\bf
  r}',t')$ in the effective potential, which includes the Hartree and
exchange-correlation contributions. It can be calculated explicitly in
frequency space according to
\begin{equation}\label{Eq:chi0}
\chi^0({\bf r},{\bf r}';\omega) = 2 \sum_{\nu,\nu'} (f_{\nu'} - f_\nu)
    \frac{\varphi_\nu({\bf r}) \varphi^*_{\nu'}({\bf r}) \varphi^*_\nu({\bf
    r}') \varphi_{\nu'}({\bf r}')}{\omega - (\epsilon_\nu - \epsilon_{\nu'}) +
    i\eta}
\end{equation}
from the static Kohn-Sham orbitals $\varphi_\nu$ and eigenvalues
$\epsilon_\nu$. The symbol $f_\nu$ indicates the Fermi occupation numbers
and $\eta$ is a positive infinitesimal. The true density-density response
function is obtained by relating it to the Kohn-Sham susceptibility through
the chain rule for functional derivatives, which may be written in the form
\begin{equation}\label{Eq:Poles}
\int\! d^3r'' \, M({\bf r},{\bf r}'';\omega) \chi({\bf r}'',{\bf r}';\omega) =
\chi^0({\bf r},{\bf r}';\omega) \;.
\end{equation}
The operator
\begin{eqnarray}
M({\bf r},{\bf r}';\omega)
&=& \delta({\bf r}-{\bf r}') - \int\! d^3r'' \, \chi^0({\bf r},{\bf
  r}'';\omega) \nonumber \\
&&\times \left( \frac{1}{|{\bf r}''-{\bf r}'|} + f^{\rm xc}({\bf r}'',{\bf
    r}';\omega) \right)
\end{eqnarray}
is related to the dielectric function and the exchange-correlation kernel is
defined as
\begin{equation}
f^{\rm xc}({\bf r},{\bf r}';t-t') = \frac{\delta V^{\rm xc}({\bf r},t)}{\delta
  n({\bf r}',t')} \;,
\end{equation}
where the functional derivative is evaluated at the unperturbed ground-state
density.

The many-body density-density response function has poles at the exact
excitation energies of the interacting electron system. On the other hand,
$\chi^0$ has poles at the excitation energies of the corresponding Kohn-Sham
system, which are in general different. Hence the singularities of $\chi$ in
Eq.\ (\ref{Eq:Poles}) must be cancelled by zeroes of the operator
$M$. Independent of the system characteristics and the nature of the excited
states, this provides a convenient starting point for calculating the
excitation spectrum. For atoms it is possible to expand all quantities in a
Laurent series around a particular Kohn-Sham energy difference $\epsilon_\nu -
\epsilon_{\nu'}$, which leads to an explicit expression for low-order
corrections to the transition energy.\cite{Petersilka1996} This procedure is
not appropriate for solids, however, because in infinite systems the Kohn-Sham
transition energies form a continuum.\cite{Mahan1990} Although the resulting
structure of $\chi^0$ can still be described in terms of effective poles,
these are located off the real axis and no longer correspond to individual
energy differences in the denominator of Eq.\ (\ref{Eq:chi0}). For the
homogeneous electron gas we therefore determine the plasmon dispersion
$\Omega(q)$ by a direct search for the zeroes of
\begin{equation}\label{Eq:MagicFunction}
M(q,\omega) = 1 - \chi^0(q,\omega) \left[ v(q) + f^{\rm xc}(q,\omega) \right]
\end{equation}
in reciprocal space, which is also the most accurate procedure. The Fourier
transform of the Coulomb potential is $v(q) = 4 \pi / q^2$, and the Kohn-Sham
susceptibility $\chi^0$ is given analytically by the dynamic Lindhard
function.\cite{Lindhard1954}

\begin{figure}[t!]
\epsfxsize=\columnwidth \centerline{\epsfbox{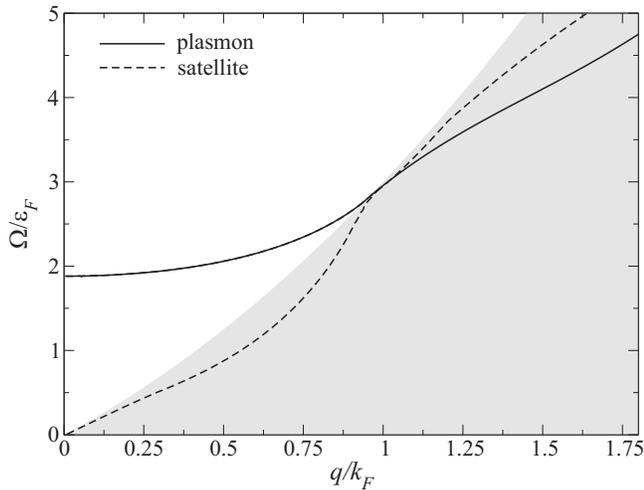}} \medskip
\caption{At the border of the electron-hole pair continuum, indicated by the
  shaded region, a crossing between the plasmon dispersion and a satellite
  excitation occurs.}\label{Fig:Satellite}
\end{figure}

In our implementation we generally calculate the zeroes of $M$ in the complex
frequency plane, which allows us to obtain both the plasmon dispersion and the
corresponding lifetimes. This approach requires the simultaneous solution of a
set of two nonlinear equations for the real and the imaginary part, for which
we employ an iterative procedure. At $q = 0$ the classical plasma frequency
$\omega_p = (4 \pi n)^{1/2}$ forms a convenient starting point, while for
finite $q$ the previously calculated solution for a smaller wavevector is
utilized. We have confirmed that the iteration is stable and convergent.
Special attention must be paid at $\Omega(q) = \frac{1}{2} q^2 + q k_F$,
however, where the structure of the excitation spectrum becomes very
complicated and a bifurcation occurs. This bifurcation, illustrated in Fig.\
\ref{Fig:Satellite} for the random-phase approximation and $r_s = 4$, where
$r_s$ denotes the Wigner-Seitz radius, is due to the crossing of the plasmon
dispersion with a satellite excitation. In order to discriminate between the
principal resonance and its satellite, we examine the spectral function
\begin{equation}
S(q,\omega) = -\frac{1}{n v(q)} \,\mbox{Im}\, \frac{1}{M(q,\omega)}
\end{equation}
on the real frequency axis, where the plasmon peak can easily be identified as
the dominant feature.

The imaginary part of $\Omega(q)$ reflects the finite lifetime of plasmons in
the homogeneous electron gas. The most important decay mechanism is scattering
into electron-hole pairs, which dominates whenever energy and momentum
conservation allow the promotion of an electron into a previously unoccupied
state above the Fermi level.\cite{Mahan1990} This region, bounded by the two
lines $\frac{1}{2} q^2 \pm q k_F$, where $k_F$ denotes the Fermi wavevector,
is shown shaded in Fig.\ \ref{Fig:Satellite}. Real solutions for the plasmon
energy, suggesting an infinite lifetime, only exist for frequency-independent
kernels outside the electron-hole pair continuum. This behavior, well known
from the random-phase approximation, constitutes a physically implausible
artefact that stems from the neglect of more intricate decay channels.

\section{Exchange-correlation kernels}\label{Sec:Kernels}

In this paper we have considered the following approximations for the
exchange-correlation kernel.

(a) In the random-phase approximation (RPA) all dynamic exchange-correlation
effects are ignored by setting
\begin{equation}
f^{\rm xc}_{\rm RPA} = 0 \;.
\end{equation}

(b) The adiabatic local-density approximation\cite{Zangwill1980} (ALDA) equals
the long-wavelength limit
\begin{equation}
f^{\rm xc}_{\rm ALDA} = \lim_{q \to 0} f^{\rm xc}_{\rm hom}(q,\omega=0)
\end{equation}
of the static exchange-correlation kernel of the homogeneous electron gas. It
is readily expressed in terms of the exchange-correlation energy per particle
$\epsilon^{\rm xc}_{\rm hom}$ as
\begin{equation}
f^{\rm xc}_{\rm ALDA} = \frac{d^2}{dn^2} \left[ n \epsilon^{\rm xc}_{\rm
    hom}(n) \right] \;.
\end{equation}
Owing to its computational simplicity, the ALDA has become the standard
approximation in time-dependent density-functional theory. It has already been
employed in calculations of the plasmon dispersion for
solids.\cite{Quong1993,Ku1999,Cazalilla2000}

(c) In their original application of time-dependent density-functional theory
to excited states, Petersilka, Gossmann, and Gross\cite{Petersilka1996} (PGG)
derived an approximate exchange-only kernel in the spirit of the optimized
effective potential method.\cite{Ulrich1995} This approach has the advantage
that the corresponding exchange potential has the proper Coulomb decay. The
kernel is constructed from the Kohn-Sham orbitals and hence only depends
implicitly on the density. Designed for small atoms, the PGG formula is
identical to the exact exchange kernel for two-electron systems, but
deviations are expected for extended systems. In particular, the frequency
dependence of the exact exchange kernel, which, in principle, can also be
calculated,\cite{Gorling1998} is ignored. In momentum space the PGG kernel is
given by\cite{Lein2000}
\begin{eqnarray}
f^{\rm xc}_{\rm PGG}(q)
&=&  -\frac{3 \pi}{10 k_F^2} \left\{ 11 + 2 Q^2  + \left( \frac{2}{Q} - 10 Q
  \right ) \ln \frac{1+Q}{|1-Q|} \right. \nonumber \\
&&+ \left. \left( 2 Q^4 - 10 Q^2 \right) \ln \left| 1 - \frac{1}{Q^2} \right|
\right\}
\end{eqnarray}
with $Q = q / 2 k_F$.

(d) Burke, Petersilka, and Gross\cite{Burke2000} (BPG) recently proposed a
hybrid formula that was shown to improve the excitation spectra of small
atoms. It combines expressions for symmetric and antisymmetric spin
orientations from different approximations in a spin density-functional
formalism. For the unpolarized homogeneous electron gas this kernel
reduces to
\begin{equation}
f^{\rm xc}_{\rm BPG}(q) = \frac{1}{2} \left[ f^{\rm xc}_{{\rm PGG},\uparrow
    \uparrow}(q) + f^{\rm xc}_{{\rm ALDA},\uparrow \downarrow} \right] \;.
\end{equation}

(e) An essentially exact parameterization of the static exchange-correlation
kernel for the homogeneous electron gas was given by Corradini, Del Sole,
Onida, and Palummo\cite{Corradini1998} (CDOP), who used the Monte Carlo
results of Moroni, Ceperley, and Senatore\cite{Moroni1995} for the static
local-field factor $G(q) = -f^{\rm xc}(q)/v(q)$. Unlike the original data, the
fit
\begin{equation}
f^{\rm xc}_{\rm CDOP}(q)= -\frac{4 \pi}{q^2} \left( C Q^2 + \frac{B Q^2}{g +
    Q^2} + \alpha Q^4 e^{-\beta Q^2} \right)
\end{equation}
with $Q = q / k_F$ is not restricted to metallic densities, because it
incorporates the known asymptotic limits for high and low densities. The
parameters $\alpha$, $\beta$, $B$, $C$, and $g$ depend on $r_s$ and are listed
in Ref.\ \onlinecite{Corradini1998}. By construction, the CDOP kernel becomes
identical to the ALDA in the long-wavelength limit.

(f) Finally we consider a parameterization of the {\em dynamic\/} local-field
factor of the homogeneous electron gas proposed by Richardson and
Ashcroft\cite{Richardson1994} (RA), including the corrections given in Ref.\
\onlinecite{Lein2000}, which stems from the summation of self-energy,
exchange, and fluctuation terms in the diagrammatic expansion of the
polarization function. It satisfies many important sum rules and includes
exact asymptotic expressions for small and large wavevectors. At intermediate
wavevectors and frequencies it provides a realistic description of the
position and magnitude of extrema, which are related to the pair distribution
function evaluated at zero separation. Because of this careful derivation we
believe the RA expression to be very close to the exact dynamic
exchange-correlation kernel of the homogeneous electron gas and give an
accurate account of the plasmon dispersion. In the absence of experimental
data we therefore use the RA results as a reference in order to assess the
performance of simpler approximations. The parameterization
\begin{equation}
f^{\rm xc}_{\rm RA}(q,\omega) = -\frac{4 \pi}{q^2} \left[ G_s(Q,U) + G_n(Q,U)
\right]
\end{equation}
with $Q = q / 2 k_F$ and $U = \omega / 4 k_F^2$ was originally given on the
imaginary frequency axis, so we use a continuation to the full complex
plane. The local-field factor $G_s$ describes screened exchange, fluctuation,
and self-energy effects in the irreducible polarizability, while $G_n$
accounts for the change in occupation numbers due to correlation. The
parameterized forms of both are listed in Ref.\ \onlinecite{Richardson1994}.

Whenever the exchange-correlation energy is needed as an input, we use the
parameterization by Perdew and Wang\cite{Perdew1992a} of the Monte Carlo data
by Ceperley and Alder.\cite{Ceperley1980} The pair-correlation function that
enters the RA kernel is taken from the same authors.\cite{Perdew1992b}

\section{Results and Discussion}\label{Sec:Results}

Before presenting our numerical results we first discuss what can be deduced
from an analytic expansion of the plasmon dispersion $\Omega(q)$. By solving
$M(q,\Omega(q)) = 0$ up to second order in $q$ we obtain the
series\cite{Dreizler1990}
\begin{equation}\label{Eq:Analytic}
\Omega(q) = \omega_p \left[ 1 + \left( \frac{9}{10 k_{\rm TF}^2} +
 \frac{f^{\rm xc}(0,\omega_p)}{8 \pi} \right) q^2 + \mbox{O}(q^4) \right] \;,
\end{equation}
where $k_{\rm TF}$ indicates the Thomas-Fermi wavevector. As long as $f^{\rm
  xc}$ does not diverge, all curves should approach the classical plasma
frequency in the long-wavelength limit. The kernel only introduces corrections
beyond the RPA in second order, where the element $f^{\rm xc}(0,\omega_p)$
appears. The ALDA evidently contains the right long-wavelength limit of the
kernel, but its neglect of the frequency dependence still introduces an error
in the parabolic term. By construction, the CDOP formula becomes identical to
the ALDA in the long-wavelength limit and thus produces the same second-order
term. Being static approximations, the PGG and BPG functionals are also
evaluated at $\omega = 0$ rather than the plasma frequency. However, they do
not approach the correct long-wavelength limit of the homogeneous electron gas
and therefore generate a different parabolic coefficient. The RA kernel, which
incorporates the full frequency dependence, is the only parameterization that
is formally exact beyond the trivial zeroth order.

\begin{figure}[b!]
\epsfxsize=\columnwidth \centerline{\epsfbox{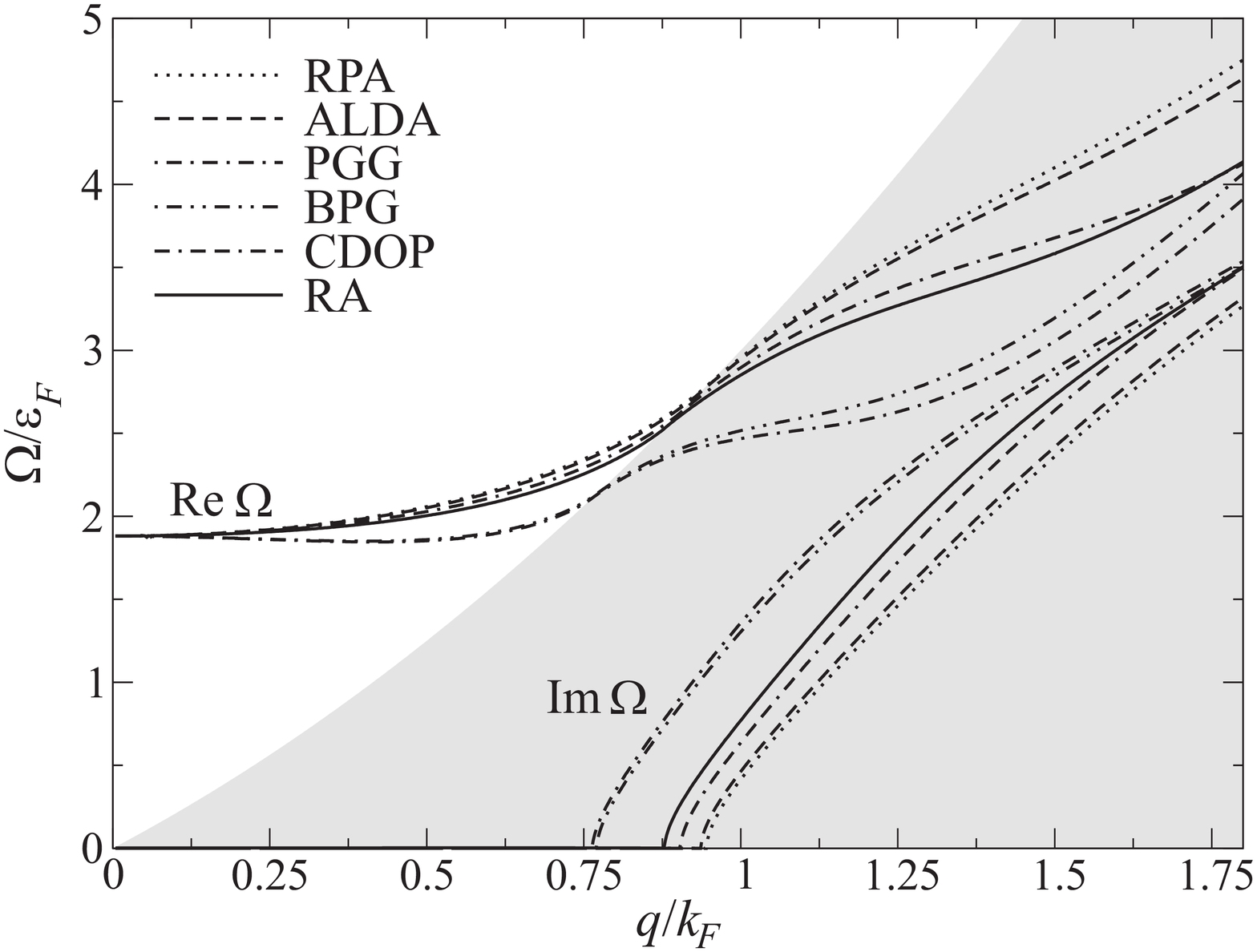}} \medskip
\caption{Plasmon dispersion for the homogeneous electron gas at $r_s = 4$,
  calculated with different approximations for the exchange-correlation
  kernel. The electron-hole pair continuum and the resulting nonzero imaginary
  part of the plasmon frequency in this regime are also
  marked.}\label{Fig:Plasmons}
\end{figure}

The numerically calculated plasmon dispersions for $r_s = 4$ are shown in
Fig.\ \ref{Fig:Plasmons}. As predicted, all curves start at the classical
plasma frequency. For small wavevectors only a small spread of the results is
observed, because the factor $9/10 k_{\rm TF}^2$ in Eq.\ (\ref{Eq:Analytic})
in general outweighs the contribution of the kernel. However, a slight
downward shift compared to the RPA is clearly visible for all nontrivial
approximations, because dynamic exchange and correlation effects combine to
lower the energy of the electron system. The ALDA and the CDOP formula produce
curves that are initially very close to the RA result we use for reference,
indicating that the neglected frequency dependence is of little consequence as
long as the correct long-wavelength limit is reproduced. This point is
emphasized by the relatively large deviation for the static PGG kernel, which
stems precisely from its incorrect behavior at $q \to 0$. The BPG curve, as
expected, lies between the ALDA and PGG results.

\begin{figure}[b!]
\epsfxsize=\columnwidth \centerline{\epsfbox{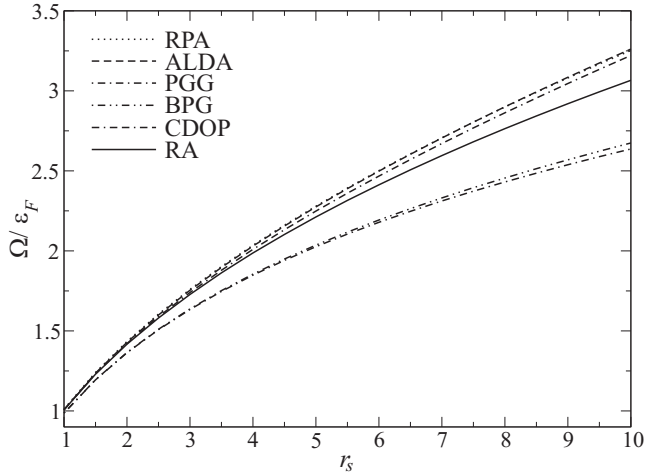}} \medskip
\caption{Behavior of the plasmon energy as a function of the density
  evaluated at $q = 0.5 q_c$, where $q_c$ indicates the critical wavevector
  corresponding to the onset of damping due to electron-hole pair excitations
  in the RPA.}\label{Fig:Smallq}
\end{figure}

To demonstrate that these observations are representative, in Fig.\
\ref{Fig:Smallq} we show the behavior of the plasmon energy over a large
density range. The curves are calculated for $q = 0.5 q_c$, where $q_c$
indicates the critical wavevector corresponding to the onset of damping due to
electron-hole pair excitations in the RPA\@. Note that the results are scaled
in units of the Fermi energy $\epsilon_F$, which is itself a function of
$r_s$. The RPA and ALDA curves are practically indistinguishable on the scale
of the figure. In the high-density limit all parameterizations tend to the RPA
result, which is the correct trend. The deviation between the RA dispersion
and the other curves increases approximately linearly with $r_s$.

\begin{figure}[t!]
\epsfxsize=\columnwidth \centerline{\epsfbox{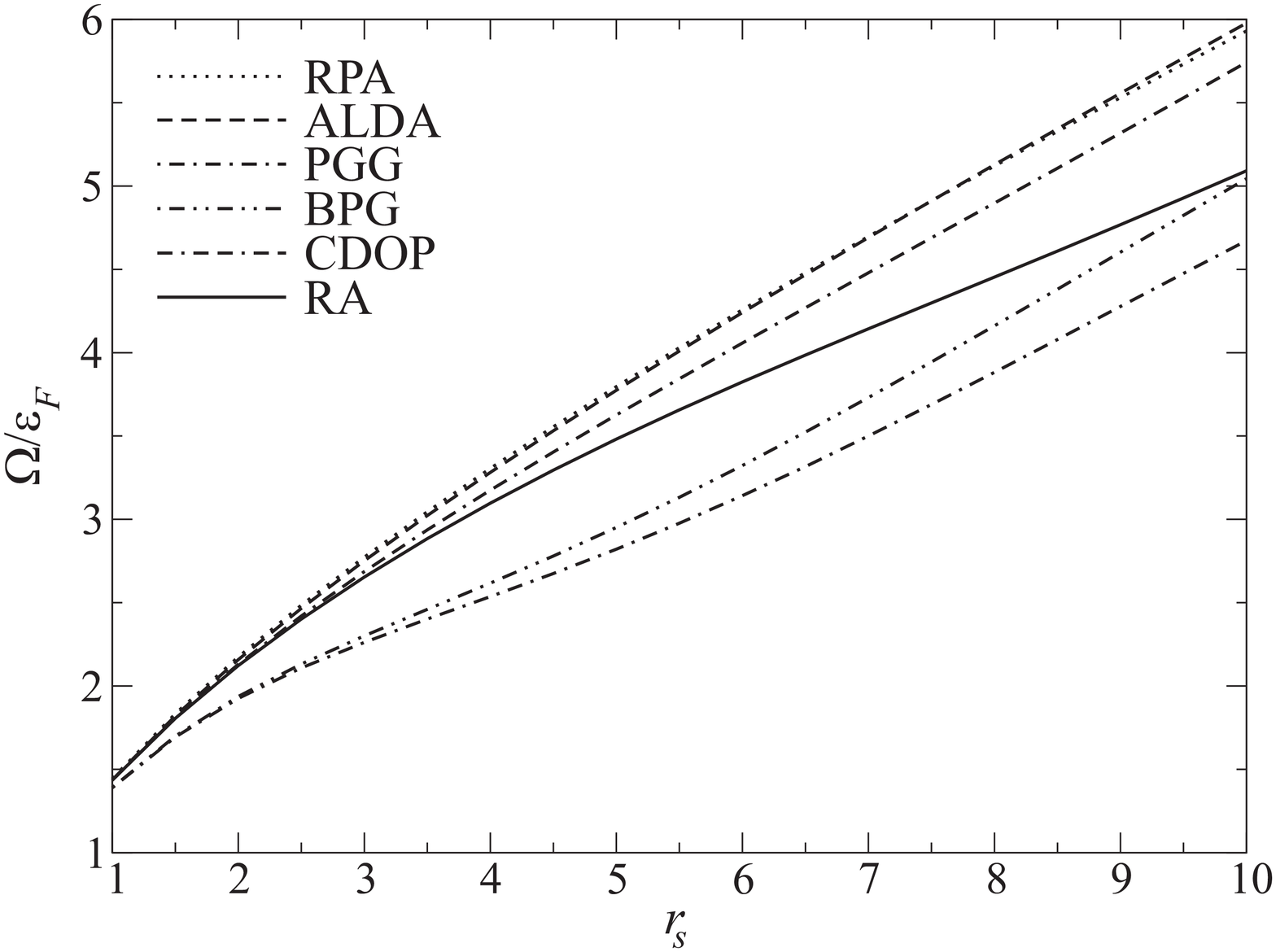}} \medskip
\caption{Behavior of the plasmon energy in the region of electron-hole pair
  excitations at $q = 1.2 q_c$.}\label{Fig:Largeq}
\end{figure}

At larger wavevectors, where the parabolic expansion (\ref{Eq:Analytic}) is no
longer valid, the differences between the considered approximations become
more pronounced. The dispersion resulting from the static ALDA kernel remains
close to the RPA at too high energies, while the CDOP result begins to deviate
slightly from the RA curve after the onset of damping in the electron-hole
pair continuum. This discrepancy must be attributed to the static nature of
the CDOP kernel. Furthermore, it can be seen that the strong downward shift of
the exchange-only PGG formula leads to an even larger error in absolute terms
than the underestimation of dynamic exchange and correlation effects in both
the ALDA and RPA\@. The hybrid BPG formula, which combines the PGG and ALDA
parameterizations, profits from a partial cancellation of errors but improves
only marginally upon PGG\@. In Fig.\ \ref{Fig:Largeq} we again show the
plasmon energy as a function of the density for $q = 1.2 q_c$.

Due to decay into electron-hole pairs in the damped regime, the plasmon energy
acquires a nonzero imaginary part, also displayed in Fig.\ \ref{Fig:Plasmons},
whose inverse is proportional to the lifetime. As a general rule we find that
all kernels yield the same quality of approximation for the imaginary part as
they do for the real part of the plasmon energy. At small wavevectors, as
discussed above, static kernels predict a vanishing imaginary part, which
corresponds to an unphysical infinite lifetime. This artefact results from
modelling $f^{\rm xc}$ as a purely real quantity by evaluating it at $\omega =
0$. In fact, the exact kernel has a finite imaginary part at nonzero
frequencies, which for small wavevectors is related to the multi-pair
component of the susceptibility according to\cite{Lipparini1994}
\begin{equation}
\mbox{Im}\, f^{\rm xc}(q,\omega) \approx -\frac{\omega^4}{\omega_p^4} [v(q)]^2
\,\mbox{Im}\, \chi^{\rm mp}(q,\omega) \;.
\end{equation}
Such multi-pair decay channels are ignored in the RPA and related schemes,
which is ultimately the reason for their qualitatively wrong behavior.
Mermin's modification of the Lindhard dielectric function avoids the problem
of infinite lifetimes,\cite{Mermin1970} but the correction based on relaxation
times is introduced in a phenomenological manner that makes it unsuitable for
{\em ab initio\/} calculations. In this study only the dynamic RA
parameterization correctly predicts a finite plasmon lifetime over the entire
frequency range. However, outside the electron-hole pair continuum the
imaginary part of the plasmon energy is several orders of magnitude smaller
than the real part and hence not discernible in the plot.

The good agreement between the static CDOP parameterization on the one hand
and the dynamic RA result on the other over a large wavevector and density
interval indicates that the frequency dependence of the kernel plays a weak
role for the plasmon dispersion. In contrast, the significant discrepancy
between static approximations like ALDA that contain the correct
long-wavelength limit and others such as PGG, which do not, suggests that a
correct parameterization of the wavevector dependence is crucial. Similar
conclusions concerning the relative importance of the frequency and wavevector
dependence were recently also reported for the correlation
energy.\cite{Lein2000} This was not entirely surprising, however, because a
frequency analysis\cite{Lein2000} reveals that the dominant contribution to
the energy in any case comes from the low-frequency limit, which is, by
design, contained correctly in all static approximations.

\begin{figure}[t!]
\epsfxsize=\columnwidth \centerline{\epsfbox{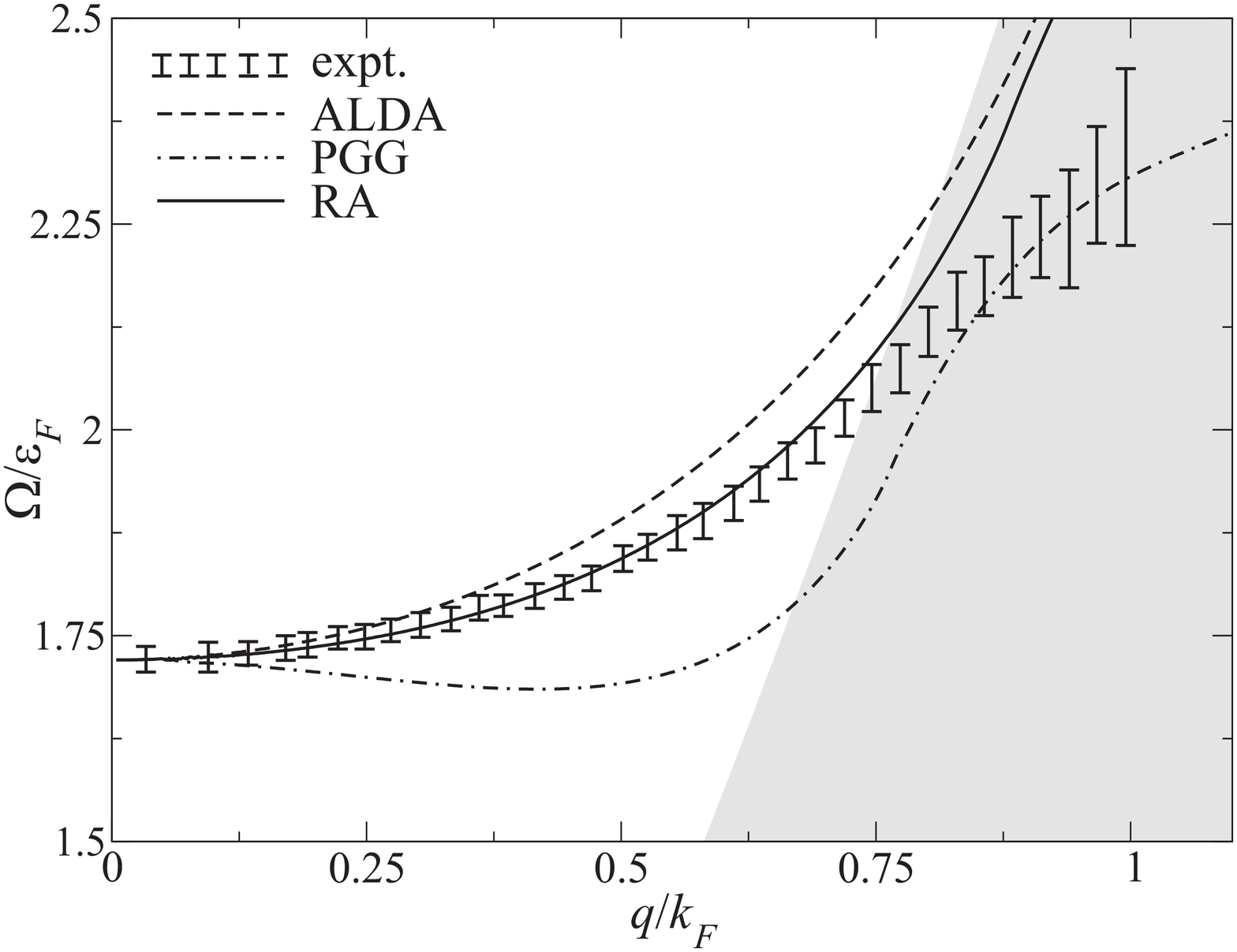}}
\caption{Calculated plasmon dispersion for $r_s = 4.0$ compared to
  experimental data for sodium from electron energy-loss spectroscopy (Ref.\
  \protect\onlinecite{vomFelde1989}). The theoretical curves have been shifted
  rigidly to the experimental value at $q = 0$ in order to account for
  core-polarization effects not included in this electron-gas
  treatment.}\label{Fig:Sodium}
\end{figure}

To emphasize the error that may arise from an inappropriate kernel, in Fig.\
\ref{Fig:Sodium} we compare theoretical results for $r_s = 4.0$ with
experimental data for sodium from electron energy-loss
spectroscopy.\cite{vomFelde1989} The theoretical curves have been shifted
rigidly to the experimental value at $q = 0$ in order to account for
core-polarization effects not included in this electron-gas
treatment.\cite{Aryasetiawan1994} At small wavevectors the RA result is in
excellent agreement with the experimental dispersion. Although the ALDA
correctly reproduces the qualitative features, its growing deviation from the
theoretical reference curve soon exceeds the experimental error bars and
becomes quite pronounced at intermediate wavevectors. This is even more
obvious in the case of PGG\@. Furthermore, for small wavevectors this
parameterization incorrectly predicts a negative dispersion that is not
observed in Na, although negative dispersion does occur in heavier alkali
metals, such as Cs.\cite{vomFelde1989} The reasons for this anomalous behavior
are still controversial.\cite{Lipparini1994,Aryasetiawan1994} Obviously, in
such situations a poor parameterization may become a serious obstacle for
theoretical interpretations. After the onset of damping due to electron-hole
pair excitations, the experimental dispersion flattens slightly. As shown in
Fig.\ \ref{Fig:Plasmons}, the theoretical results exhibit the same effect,
but the unshifted RA and ALDA curves only cross the border of the damped
regime at larger critical wavevectors. Hence in Fig.\ \ref{Fig:Sodium}
quantitative agreement cannot be expected for large wavevectors due to the
different physical situations. The PGG curve, on the other hand, lies below
the RA result and consequently enters the damped regime at a smaller critical
wavevector, but the good agreement with the experimental dispersion for large
wavevectors in Fig.\ \ref{Fig:Sodium} is clearly fortutious.

\section{Summary}\label{Sec:Summary}

In this paper we have tested several common approximations for the
exchange-correlation kernel by examining the plasmon dispersion of the
homogeneous electron gas. First of all, we have found that the influence of
the kernel is indeed significant, giving rise to large differences between the
calculated dispersion curves. The ALDA performs reasonably well, although it
underestimates dynamic exchange-correlation effects embodied in the kernel and
improves only little upon the RPA\@. A better quantitative scheme is therefore
desirable. In this respect our results, in particular the good agreement
between the dynamic RA parameterization and the static CDOP kernel, both of
which are presumed to be very accurate, suggest that the neglect of the
frequency dependence is of little consequence, giving rise to small deviations
only at large wavevectors in the electron-hole pair continuum. The error
of the ALDA thus stems largely from its local nature, and extensions should
focus on a better description of the wavevector dependence. The challenge of
this task is illustrated by the fact that some of the explicitly nonlocal
parameterizations we considered, notably PGG, actually lead to worse results
although they are known to improve excitation spectra in small atoms. This
apparent paradox may be understood by the significance of the long-wavelength
limit for the homogeneous electron gas, which determines the leading order of
the plasmon dispersion and is contained correctly in the ALDA but not in the
PGG kernel. In localized systems such as atoms, on the other hand, the
long-wavelength limit is less relevant, whereas dynamic exchange effects
contained in the PGG kernel may play an important role. This lack of
transferability should encourage specific approximations for solids. The CDOP
kernel, which derives from the homogeneous electron gas, seems a step into the
right direction, although its performance for real materials has not been
fully explored yet. Finally, we demonstrated that exchange-correlation kernels
optimized for small atoms may lead to quantitative, and occasionally
qualitative, deviations in the plasmon dispersion for solids that
significantly exceed the corresponding experimental error bars and may affect
theoretical interpretations.

\acknowledgements

This work was funded in part by the EU through the NANOPHASE Research Training
Network (Contract No.\ HPRN-CT-2000-00167). We thank M.\ Fuchs for a careful
reading of the manuscript.


\begin{references}
\bibitem[*]{email} Electronic address: tatar@fhi-berlin.mpg.de
\bibitem{Hohenberg1964} P. Hohenberg and W. Kohn, Phys. Rev. {\bf 136}, B864
  (1964); W. Kohn and L. J. Sham, Phys. Rev. {\bf 140}, A1133 (1965).
\bibitem{Dreizler1990} R. M. Dreizler and E. K. U. Gross, {\em Density
    Functional Theory\/} (Springer, Berlin, Heidelberg, 1990).
\bibitem{Runge1984} E. Runge and E. K. U. Gross, Phys. Rev. Lett. {\bf 52},
  997 (1984).
\bibitem{Petersilka1996} M. Petersilka, U. J. Gossmann, and E. K. U. Gross,
  Phys. Rev. Lett. {\bf 76}, 1212 (1996).
\bibitem{Aryasetiawan1998} F. Aryasetiawan and O. Gunnarsson,
  Rep. Prog. Phys. {\bf 61}, 237 (1998).
\bibitem{Casida1998a} M. E. Casida, K. C. Casida, D. R. Salahub,
  Int. J. Quantum Chem. {\bf 70}, 933 (1998).
\bibitem{Hirata1999} S. Hirata and M. Head-Gordon, Chem. Phys. Lett. {\bf
    302}, 375 (1999).
\bibitem{Sundholm1999} D. Sundholm, Chem. Phys. Lett. {\bf 302}, 480 (1999).
\bibitem{Olevano1999} V. Olevano, M. Palummo, G. Onida, and R. Del Sole,
  Phys. Rev. B {\bf 60}, 14~224 (1999).
\bibitem{Kootstra2000} F. Kootstra, P. L. de Boeij, and J. G. Snijders,
  J. Chem. Phys. {\bf 112}, 6517 (2000).
\bibitem{Quong1993} A. A. Quong and A. G. Eguiluz, Phys. Rev. Lett. {\bf 70},
  3955 (1993); A. Fleszar, A. A. Quong, and A. G. Eguiluz,
  Phys. Rev. Lett. {\bf 74}, 590 (1995).
\bibitem{Ku1999} W. Ku and A. G. Eguiluz, Phys. Rev. Lett. {\bf 82}, 2350
  (1999); A. G. Eguiluz, W. Ku, and J. M. Sullivan, J. Phys. Chem. Solids {\bf
    61}, 383 (2000).
\bibitem{Cazalilla2000} M. A. Cazalilla, J. S. Dolado, A. Rubio, and
  P. M. Echenique, Phys. Rev. B {\bf 61}, 8033 (2000).
\bibitem{Umrigar1994} C. J. Umrigar and X. Gonze, Phys. Rev. A {\bf 50}, 3827
  (1994).
\bibitem{Casida1998b} M. E. Casida, C. Jamorski, K. C. Casida, and
  D. R. Salahub, J. Chem. Phys. {\bf 108}, 4439 (1998).
\bibitem{Burke2000} K. Burke, M. Petersilka, and E. K. U. Gross, in {\em
    Recent Advances in Density Functional Methods}, edited by P. Fantucci and
  A. Bencini (World Scientific, Singapore, in press), Vol. III.
\bibitem{Corradini1998} M. Corradini, R. Del Sole, G. Onida, and M. Palummo,
  Phys. Rev. B {\bf 57}, 14~569 (1998).
\bibitem{Lein2000} M. Lein, E. K. U. Gross, and J. Perdew, Phys. Rev. B {\bf
    61}, 13~431 (2000).
\bibitem{Kluner1998} T. Kl\"uner, H.-J. Freund, V. Staemmler, and R. Kosloff,
  Phys. Rev. Lett. {\bf 80}, 5208 (1998). 
\bibitem{Mahan1990} G. D. Mahan, {\em Many-Particle Physics\/} (Plenum, New
  York, 1990).
\bibitem{Lindhard1954} J. Lindhard,
  K. Dan. Vidensk. Selsk. Mat. Fys. Medd. {\bf 28}, no. 8 (1954).
\bibitem{Zangwill1980} A. Zangwill and P. Soven, Phys. Rev. Lett. {\bf 45},
  204 (1980).
\bibitem{Ulrich1995} C. A. Ullrich, U. J. Gossmann, and E. K. U. Gross,
  Phys. Rev. Lett. {\bf 74}, 872 (1995).
\bibitem{Gorling1998} A. G\"orling, Phys. Rev. A {\bf 57}, 3433 (1998).
\bibitem{Moroni1995} S. Moroni, D. M. Ceperley, and G. Senatore,
  Phys. Rev. Lett. {\bf 75}, 689 (1995).
\bibitem{Richardson1994} C. F. Richardson and N. W. Ashcroft, Phys. Rev. B
  {\bf 50}, 8170 (1994).
\bibitem{Perdew1992a} J. P. Perdew and Y. Wang, Phys. Rev. B {\bf 45}, 13~244
  (1992).
\bibitem{Ceperley1980} D. M. Ceperley and B. J. Alder, Phys. Rev. Lett. {\bf
    45}, 566 (1980).
\bibitem{Perdew1992b} J. P. Perdew and Y. Wang, Phys. Rev. B {\bf 46}, 12~947
  (1992); {\bf 56}, 7018(E) (1997).
\bibitem{Lipparini1994} E. Lipparini, S. Stringari, and K. Takayanagi,
  J. Phys.: Condens. Matter {\bf 6}, 2025 (1994). 
\bibitem{Mermin1970} N. D. Mermin, Phys. Rev. B {\bf 1}, 2362 (1970).
\bibitem{vomFelde1989} A. vom Felde, J. Spr\"osser-Prou, and J. Fink,
  Phys. Rev. B {\bf 40}, 10~181 (1989).
\bibitem{Aryasetiawan1994} F. Aryasetiawan and K. Karlsson,
  Phys. Rev. Lett. {\bf 73}, 1679 (1994). 
\end{references}
\end{document}